\documentclass[preprint,showpacs,preprintnumbers,amsmath,amssymb,prb]{revtex4-1} 

\usepackage{graphicx}% Include figure files
\usepackage{dcolumn}% Align table columns on decimal point
\usepackage{bm}% bold math
\usepackage{color} %If we use color, figure can be not placed properly. 
\begin{document}
\title{Motional Narrowing under Markovian and Non-Markovian Hopping Transitions 
in Inhomogeneous Broadened Absorption Line Shape}
\author{
Kazuhiko Seki
}
%\email{k-seki@aist.go.jp}
\affiliation{Nanomaterials Research Institute(NMRI), 
National Institute of Advanced Industrial Science and Technology (AIST)\\
AIST Tsukuba Central 5, Higashi 1-1-1, Tsukuba, Ibaraki 305-8565, Japan 
}
\author{Kazuhiro Marumoto}
%\altaffiliation[Permanent address:]{
\affiliation{Division of Materials Science, University of Tsukuba, Tsukuba, Ibaraki 305-8573, Japan
}
%\preprint{}
\begin{abstract}
%%% revised 
Inspired by recent experiments showing a minimum of 
electron paramagnetic resonance (ESR/EPR) line width as a function of inverse temperature,
%%% revised 
we studied the motional narrowing effect by considering a combined model of carrier transitions and static dispersion of the angular frequency giving rise to an inhomogeneous broadening in the spectrum. The dispersion of the angular frequency results from the distribution of the local field. The transition between the sites under inhomogeneous static local field induces adiabatic relaxation of the spin. We also considered the on-site 
%%% revised 
inherent (nonadiabatic) 
%%% revised 
relaxation of the spin. We obtained the exact solution of the spin correlation function by explicitly considering transitions between two sites for both Markovian and non-Markovian transition processes. The absorption line shape is expressed in terms of the Voigt function, which is a convolution of a Gaussian function and a Lorentzian function. Using the known properties of the Voigt function, we discuss the correlation between the change in the full-width at half-maximum and the change in line shape, both of which are induced by motional narrowing. 
By assuming thermal activation processes for both the hopping transition and the on-site %%% revised 
inherent 
%%% revised 
relaxation, 
we show that the minimum of the width appears as a function of inverse temperature as observed experimentally in organic 
%%% revised 
materials.
%%% revised
Contrary to the general belief, we also show that the narrowing of the Gaussian line shape under a local random field did not necessarily lead to a Lorentzian line shape in particular under the presence of heavy tail property in the waiting time distribution of 
hopping transitions. 
\end{abstract}
%\PACS 78.55.Qr \sep 72.20.Ee \sep 05.40.a

\maketitle
\section{Introduction}
\label{sec:I}
Electron paramagnetic resonance (ESR/EPR) is a powerful tool in the study of charge transport in organic electronic 
%%% revised 
materials 
%%% revised
because ESR selectively detects unpaired conducting electrons and is insensitive to bounded paired electrons.
The ESR technique has been applied to reveal molecular orientation with respect to the substrate, carrier doping effects, delocalized carrier states, and carrier transport properties.\cite{Namatame_17, Marumoto_JPSJ,Marumoto_06,Tanaka_06,Marumoto_11,Matsui_10,Matsui_12,Matsui_08}
Recently, motional narrowing was observed in organic 
%%% revised 
materials 
%%% revised
such as pentacene and rubrene by varying temperature; \cite{Matsui_08,Matsui_12,Matsui_10,Marumoto_11,Tanaka_06}
%%% revised 
the width of the line shape shows a minimum as a function inverse temperature. \cite{Matsui_08,Matsui_12,Matsui_10}
%%% revised 
In some instances, this motional narrowing was associated with an absorption spectral change from a Gaussian absorption line shape to a Lorentzian line shape but, interestingly, it was sometimes observed without an apparent change in the line shape.\cite{Matsui_08,Matsui_12, Matsui_10,Marumoto_11,Tanaka_06}
The motional narrowing has been investigated theoretically, mainly in the time domain, including the analytical work of Kubo and Anderson;\cite{Kubo_54,Kubo_Tomita,Anderson_54,Kubo_adv} the asymptotic time dependence of the spin correlation function has been investigated but the complete kinetics has received less attention. 
In ESR, motional narrowing is measured using the spectrum of the first derivative of the absorption line shape. 
Although motional narrowing has been known for decades,\cite{slichter,BPP} its effect on the absorption line shape has not been fully determined compared with the magnetic relaxation in the time domain. 

Qualitative features of motional narrowing can be understood by analyzing the Kubo--Anderson model.\cite{Kubo_54,Kubo_Tomita,Anderson_54,Kubo_adv}
In this model, the influence of the local field fluctuation is assumed to involve Gaussian and Markovian random processes.\cite{Kubo_54,Kubo_Tomita,Anderson_54,Kubo_adv}
The fluctuating field is thought to reflect the change in the magnetic field imposed on the spin by transport in an inhomogeneous medium. 
In the limit of slow decay times of the local field fluctuations, a Gaussian decay of the spin correlation was obtained, whereas in the limit of fast decay, an exponential decay of the spin correlation was obtained. 
In this stochastic description, the change in the decay of the spin correlation function corresponds to motional narrowing of the absorption line shape. 
The Gaussian decay and the exponential decay lead to a Gaussian and a Lorentzian absorption line shape, respectively. 
Qualitatively different line shapes are expected. 
Nevertheless, we find the more reserved statement "narrowing process of the Gaussian spectrum in the presence of Markovian motion" in the original paper by Kubo;\cite{Kubo_54} the nature of the narrowed spectrum was not specified.

It has been recognized that the fluctuating field does not obey Gaussian and Markovian random processes when field fluctuations originate from the movement of a magnetic carrier in inhomogeneous magnetic-field environments.\cite{Czech_84,Czech_86,Mitra_91}
In these studies, an exponential decay is obtained as the long-time asymptotic decay of the spin correlation regardless of the dimensionality of the kinematic space. 
Although the asymptotic decay was exponential as predicted assuming both Gaussian and Markovian random processes, the relaxation rate constant differed from the results obtained under this assumption in particular for low dimensions. 
These studies pointed out the importance of taking into account the actual movement of spin carriers in inhomogeneous local-field environments. 
Moreover, absorption line shapes have received less attention in theoretical work. 

In this paper, we take into account explicitly the motion of the spin carrier under random local fields. 
For simplicity, we consider two-site transition processes. 
The local field at each site obeys a Gaussian distribution. 
We obtain the exact analytical expression for the spin correlation function for both Markovian and non-Markovian transition processes. 
Using the exact expression for Markovian transition processes, 
the spin correlation function is expressed by the function similar to that given by the familiar expression of the Kubo--Anderson model.\cite{Kubo_Tomita,Kubo_54} 
The slight difference manifests itself when the transition rate between sites is large. 
Therefore, the narrowed absorption line shape from a fast transition between sites may differ from the Lorentzian line shape. 
The absorption line shape is expressed in terms of the Voigt function, which is a convolution of Gaussian and Lorentzian functions. 
Using the known properties of Voigt function,\cite{OLIVERO_77} we discuss the correlation between the change in the full-width at half-maximum (FWHM) and the change in line shape, both of which are induced by motional narrowing. 
To confirm the above-mentioned results, we also present the spectrum of the first derivative, which is relevant to ESR measurements. 
%%% revised 
By using the above mentioned results, 
we show that the minimum of the width appears as a function of inverse temperature as observed experimentally in organic materials.
%%% revised 
We also show that the narrowing of the Gaussian line shape does not necessarily lead to a Lorentzian line shape under a random local field.

%%%
\section{Theory}
\label{sec:II}
\begin{figure}[h]
  \begin{center}
    \includegraphics[width=0.3\textwidth]{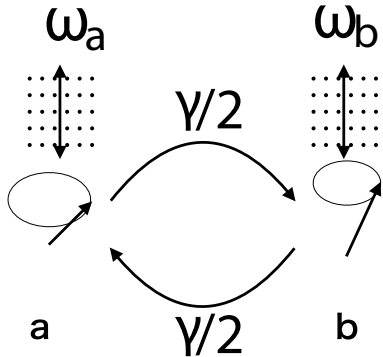}
  \end{center}
  \caption{Schematic picture of the Kubo--Anderson--Kitahara model is shown.  
  %%% revised  
The model comprises a spin executing transitions between 
a pair of  statistically equivalent sites 
labeled by a and b. We consider a large number of the pairs and study the ensemble average of spins. 
  %%% revised 
  The local field at each site obeys a Gaussian distribution. As a result of local field dispersion, the angular frequency of the spin at each site obeys a Gaussian distribution centered on the Larmor angular frequency. The carrier transition rate between the sites is given  by $\gamma/2$. We also include the spin relaxation at each site with the rate $\gamma_s$.}
  \label{fig:KAK}
\end{figure}
%%% revised   
Introduced by Kitahara,\cite{KK_87,KOK_87, Kitahara_94} the model comprises a spin executing transitions between  
a pair of  statistically equivalent sites 
labeled by a and b in Fig.~\ref{fig:KAK}.  
We consider a large number of the pairs and study relaxation of spins by 
transitions between two sites with local field at each site. 
We study the ensemble average of spins when  
a large number of sites are exposed to different local field environments. 
The local field at each site obeys a Gaussian distribution (Figure~\ref{fig:KAK}). 
%%% revised  

Adiabatic relaxation occurs via spin transitions under an inhomogeneous local magnetic field; 
%%% revised  
the time correlation of spins results from recursive nature of spin transitions even though 
local fields are spatially uncorrelated, {\it i.e.}  spin correlations are generated by 
returning to the originally occupied site where  
the time independent and random magnetic field is assigned. 
%%% revised  
The ensemble average of the magnetic moments of the carriers precesses under different local fields and decays by Markovian transitions. 
In addition, we consider on-site 
%%% revised 
inherent (nonadiabatic) 
%%% revised
relaxation characterized by the decay rate constant denoted by $\gamma_{s}$. 
Within linear response theory, the width of the absorption lines is characterized by the transverse relaxation time.\cite{Toda_92}
Therefore, we consider transverse relaxation as in the Kubo--Anderson model,\cite{Kubo_54, Kubo_adv,Anderson_53,Anderson_54} where the external magnetic field is applied in the z-direction in the Cartesian coordinate system that is erected. 
For the spin labeled by the site $i=a \mbox{ or } b$, we introduce $M_{i}=M_{i,x}+iM_{i,y}$ in this coordinate system. 
We denote by $\omega_i$ the angular frequency of the spin rotating around a fixed magnetic field in the z-direction. 
We assume that a Gaussian distribution of the magnetic fields at all sites; that is,
$\omega_a$ and $\omega_b$ are statistically independent and obey Gaussian distributions. 
The mean of $\omega_i$ is zero and the variance is denoted by $\Delta_0^2$. 
The transition rate between the sites is denoted by $\gamma$. 
When a static magnetic field is applied along the z-axis, the magnetization $M_i^{(z)}(t)$ undergoes a precession around the field direction with Larmor angular frequency $\omega_0$. 
Because we are interested in the dispersion of the magnetization around the coherent motion with the Larmor angular frequency, we study the magnetization given by $M_i (t)=M_i^{(z)}(t)\exp(i \omega_0 t)$. 
The kinetic equation for the spins is written 
\begin{align}
\frac{d}{dt} M_a(t)=i \omega_a M_a(t) - \gamma_{s} M_a(t)-
\frac{\gamma}{2}\left( M_a(t)- M_b(t) \right) 
\label{eq:spina}\\
\frac{d}{dt} M_b(t)=i \omega_b M_b(t) - \gamma_{s} M_b(t)-
\frac{\gamma}{2}\left( M_b(t)- M_a(t) \right),
\label{eq:spinb}
\end{align}
where we consider the initial conditions at equilibrium given by $M_a(0)=M_b(0)=M_0/2$. 

We calculate the spin correlation averaged over the distribution of the local Larmor angular frequency and the initial condition 
\begin{align}
\langle M(t)/M(0) \rangle=\phi(t)
\label{eq:spincorr}
\end{align}
and the line-shape function using the Wiener--Khinchin theorem, 
\begin{align}
I(\omega)=\frac{1}{2\pi} \int_{-\infty}^\infty dt \phi(t) \exp(-i \omega t). 
\label{eq:spinspe}
\end{align}
%%% modified 
The term $\langle \cdots \rangle$ in Eq.~(\ref{eq:spincorr}) can be 
expressed as 
$1/(2 \pi \Delta_0^2)\int_{-\infty}^\infty d\,\omega_a \int_{-\infty}^\infty d\,\omega_b
\exp \left[-(\omega_a ^2+\omega_b^2)/(2\Delta_0^2)\right] \cdots$. 
%%% modified 
The power spectrum gives 
the absorbance of the microwave radiation 
multiplied by $\omega_0$ if $\omega_0 \gg \omega_i$ is satisfied for the majority of the distribution of $\omega_i$.\cite{Kubo_Tomita}
%%% changed
The ESR absorption  
%%% changed
is essentially proportional to the power spectrum of the autocorrelation function of $M(t)$.

%%% revised
As shown in the Appendix A,  
%%% revised 
the spin correlation function becomes
\begin{align}
\phi(t) =\exp \left( - \frac{1}{4}\Delta_0^2 t^2 - 2 \gamma_{s} t \right)
\sum_{m=0}^\infty \frac{(-1)^m}{m!} \left(\frac{\Delta_0 t}{2} \right)^{2m} 
{\cal M} \left(m; 2m+1; -\gamma t \right).
\label{eq:spincorr_e}
\end{align}

For a slow transition rate expressed by $\gamma t \ll 1$, taking the limit simplifies Eq.~(\ref{eq:spincorr_e}) to
\begin{align}
\phi_{s}(t) \approx \exp \left( - \frac{1}{2}\Delta_0^2 t^2 - 2 \gamma_{s} t \right), 
\label{eq:spincorr_s}
\end{align}
where $\sum_{m=0}^\infty (-1)^m \left(\Delta_0 t/2 \right)^{2m} /m!=\exp \left( - \Delta_0^2 t^2 /4\right)
$ is used. 
In contrast, for a fast transition rate expressed by $\gamma t \gg 1$, taking the limit reduces Eq.~(\ref{eq:spincorr_e}) to 
\begin{align}
\phi_{f}(t) \approx \exp \left( - \frac{1}{4}\Delta_0^2 t^2 - 2 \gamma_{s} t \right),  
\label{eq:spincorr_f}
\end{align}
where the asymptotic expansion given by ${\cal M} \left(m; 2m+1; -\gamma t \right)\approx \Gamma(2m+1)/[m! (\gamma t)^m]$ 
is used. 
This equation indicates a slowing down of the spin relaxation in the fast-transition-rate limit because the factor given by $(1/2) \Delta_0^2$ in Eq.~(\ref{eq:spincorr_s}) changes to $(1/4) \Delta_0^2$ in Eq.~(\ref{eq:spincorr_f}).\cite{Kitahara_94,KK_87,KOK_87}
Correspondingly, the absorption spectral line is narrowed.\cite{Kitahara_94,KK_87,KOK_87} 
The result of Eq.~(\ref{eq:spincorr_f}) can be interpreted by averaging $\exp(i \omega_m t)$ using the distribution of $\omega_m$ given by Eq.~(\ref{eq:2018_2_6_2_1}).
Here, $\omega_m$ denotes the mean angular frequency between the sites. 
The result implies that the spin precesses at the mean angular frequency between the sites in the fast-transition-rate limit. 
The limit was also found for quantum tunneling.\cite{KK_87,KOK_87} 
By further calculating an ensemble average with respect to the angular frequency, the relaxation becomes a Gaussian function. 
%%% revised 
In the fast-transition-rate limit, 
the spin precesses at the mean angular frequency between the sites 
due to recursive nature of transitions. 
Recursive nature of transitions might still remain 
for spatially extended systems, in particular in low-dimensional systems; 
the result could depend on the lattice structure.  
We also reserve the possibility that the spin relaxation function given by Eq. (\ref{eq:spincorr_f}) could 
result from confinement effect; 
the recursive feature of transitions could be enhanced by confinement in a finite volume. 
%%% revised 

A correction to Eq.~(\ref{eq:spincorr_f}) may be calculated from Eq.~(\ref{eq:spincorr_e}), the result being
\begin{align}
\phi_f (t) \approx \exp \left( - \frac{1}{4}\Delta_0^2 t^2 - 2 \gamma_{s} t \right)
\left[1-\frac{1}{2} \left( \frac{\Delta_0}{\gamma} \right)^2 
\left(\gamma t -1 + e^{-\gamma t} \right) \right], 
\label{eq:spincorr_f1}
\end{align}
where the summation in Eq.~(\ref{eq:spincorr_e}) is taken up to $m=1$;  
${\cal M} \left(0; 1; x\right)=1$ and ${\cal M} \left(1; 3; -x\right)=2[x-1+\exp(-x)]/x^2$ are introduced. 
Using the approximation given by
\begin{align}
1-\frac{1}{2} \left( \frac{\Delta_0}{\gamma} \right)^2 
\left(\gamma t -1 + e^{-\gamma t} \right) \approx 1- \left(\frac{1}{2} \Delta_0 t\right)^2,
\label{eq:spincorr_f2}
\end{align}
we expand Eq.~(\ref{eq:spincorr_f1}) and find $1 - \frac{1}{2}\Delta_0^2 t^2$. 
This expression coincides with the short time expansion of Eq.~(\ref{eq:spincorr_s}). 
On the basis of this observation, Eq.~(\ref{eq:spincorr_f1}) is modified to 
\begin{align}
\phi (t) \approx \exp \left[ - \frac{1}{4}\Delta_0^2 t^2
- 2 \gamma_{s} t -\frac{1}{2} \left( \frac{\Delta_0}{\gamma} \right)^2 
\left(\gamma t -1 + e^{-\gamma t} \right) \right], 
\label{eq:spincorr_m}
\end{align}
which, given in terms of $\gamma$, makes the expression applicable to longer time durations regardless of the limit. 
The result of Eq.~(\ref{eq:spincorr_m}) is close to the numerically exact result [Figure~\ref{fig:corr_spec}(a)]. 
In the absence of the $-(1/4)\Delta_0^2 t^2$--term and $\gamma_s$, the right-hand side of Eq.~(\ref{eq:spincorr_m}) is a well-known function describing the narrowing of the Gaussian profile through increasing $\gamma$.\cite{Kubo_Tomita,Kubo_54,Anderson_54}
In the absence of the $-(1/4)\Delta_0^2 t^2$ term and $\gamma_st$, the functional form was regarded as the basis for understanding motional narrowing of line shapes. 
Here, Eq.~(\ref{eq:spincorr_m}) including the $-(1/4)\Delta_0^2 t^2$ term was obtained in an approximation by considering a combined model of carrier transitions and a static dispersion of the angular frequency that gave rise to inhomogeneous broadening in the spectrum. 
The exact result for the model is given by Eq.~(\ref{eq:spincorr_e}), which along with
Eq.~(\ref{eq:spincorr_m}), constitute in part the main results of this paper. 

%%%
\section{Spectrum}
\label{sec:III}

The spectrum is obtained numerically by substituting Eq.~(\ref{eq:spincorr_e}) into Eq.~(\ref{eq:spinspe}); we also obtained analytical expressions. 
We have
\begin{align}
I(\omega)\approx V(\omega; \Delta_0,2\gamma_{s}), 
\label{eq:spectrum_s}
\end{align}
in the limit of small value of $\gamma$, and 
\begin{align}
I(\omega)\approx V(\omega; \Delta_0/\sqrt{2},2\gamma_{s})
\label{eq:spectrum_f}
\end{align}
in the limit of large value of $\gamma$. 
In both limits, the spectrum is expressed by the Voigt function,\cite{NIST}
\begin{align}
V(x; \sigma,r)=\mbox{Re}\left[\exp\left(-\frac{(x-ir)^2}{2\sigma^2} \right)
\mbox{erfc} \left(\frac{(ix+r)}{\sqrt{2}\, \sigma}\right) \right]/\sqrt{2\pi \sigma^2}.
\label{eq:Voigt_f}
\end{align}
Denoting the complex conjugate of a function $f$ by $f^*$, we have $Re(fg)=Re(f^*g^*)$. 
Therefore, Eq.~(\ref{eq:Voigt_f}) can be rewritten as
$V(x; \sigma,r)=\mbox{Re}[\exp(-z^2)\mbox{erfc}(-i z)]/\sqrt{2\pi \sigma^2}$, where $z=(x+ir)/(\sqrt{2}\, \sigma)$. 
$w(z)=\exp(-z^2)\mbox{erfc}(-i z)$ is called the Faddeeva function.\cite{NIST}
As a convolution of a Gaussian profile $P_G(x)$ and a Lorentzian profile $P_L(x)$, the Voigt profile is
\begin{align} 
V(x; \sigma,r)=\int_{-\infty}^\infty dy P_G(y;\sigma)P_L(x-y;r),
\label{eq:Voigt}
\end{align}
where the Gaussian and Lorentzian profiles are given by 
\begin{align}
P_G(x;\sigma)&=\frac{1}{\sqrt{2\pi \sigma^2}}=\exp\left(-\frac{x^2}{2\sigma^2}\right),
\label{eq:G}\\
P_L(x;r)&=\frac{r}{\pi(x^2+r^2)},
\label{eq:L}
\end{align}
respectively.
The Voigt function describes a combined broadening of the Gaussian and Lorentzian line shapes. 

\begin{figure}[h]
\begin{center}
    \includegraphics[width=0.5\textwidth]{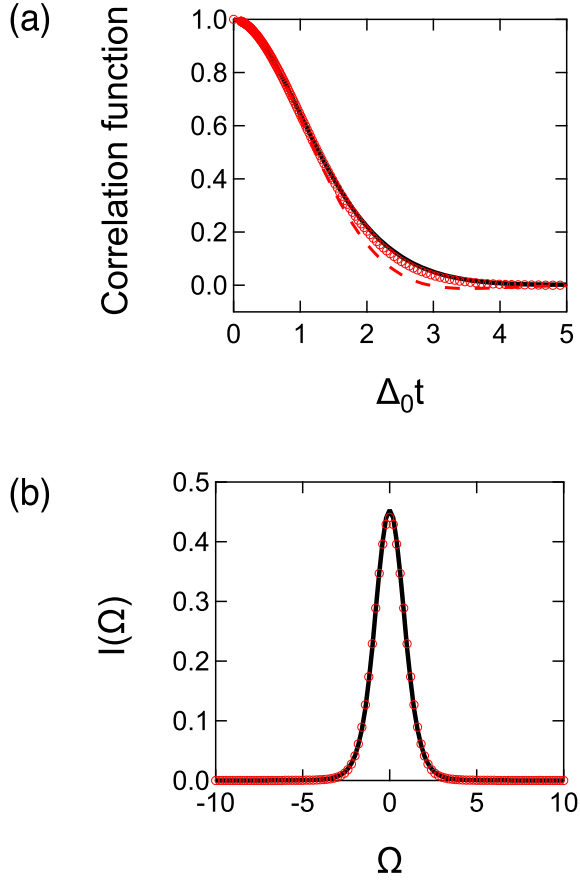}
\end{center}
\caption{(Color online) (a) Spin correlation function is shown as a function of time. (b) The power spectrum is shown as a function of angular frequency. In both plots, we set $\Gamma_{s}=0.01$ and $\Gamma=1$; the solid lines represent the numerically exact results. In (a), the red circles represent the results obtained from Eq.~(\ref{eq:spincorr_m}); the red dashed line represents the approximate solution obtained from Eq.~(\ref{eq:spincorr_f1}). In (b), the red circles represent approximate results obtained from Eq.~(\ref{eq:width_Voigt_f_sum}) by summing the first four terms.}
\label{fig:corr_spec}
\end{figure}

For the numerical calculation, we introduce dimensionless parameters using $\Delta_0$ as 
\begin{align}
\Omega=\frac{\omega}{\Delta_0},\Gamma_{s}=\frac{2\gamma_{s}}{\Delta_0} \mbox{ and } \Gamma=\frac{\gamma}{\Delta_0}. 
\label{eq:dimensionless}
\end{align}
We systematically vary these parameters to study the absorption line shape. 
Dimensionless parameters are also used to obtain an approximate expression valid for a wide range of $\gamma$. 
We first express Eq.~(\ref{eq:spincorr_m}) using $\tau=\Delta_0t$ as 
\begin{align}
\phi (\tau) \approx \sum_{j=0}^\infty \frac{1}{j!} \left(-\frac{1}{2\Gamma^2} \right)^j
\exp \left[ - \frac{1}{4}\tau^2 - \left(\Gamma_{s} 
+\frac{1}{2\Gamma} + \Gamma j\right) \tau + \frac{1}{2\Gamma^2}\right]. 
\label{eq:2018_2_16_8}
\end{align}
By substituting Eq.~(\ref{eq:2018_2_16_8}) into Eq.~(\ref{eq:spinspe}), we obtain the approximate expression 
\begin{align}
I(\Omega) =\sum_{j=0}^\infty \frac{1}{j!}
\left(-\frac{1}{2\Gamma^2} \right)^j \exp\left(\frac{1}{2\Gamma^2} \right)
V\left(\Omega; \frac{1}{\sqrt{2}},\Gamma_{s}+\frac{1}{2\Gamma}+\Gamma j \right). 
\label{eq:width_Voigt_f_sum}
\end{align}
The power spectra were expressed in terms of the weighted sum of the Voigt distribution, where the weight is expressed as a power of $-(1/2\Gamma^2)$. 
As shall be shown later [Figure~\ref{fig:corr_spec}(b)], the approximate expression reproduces the numerically exact result by summing the first four terms as long as the condition $\Gamma>1/ \sqrt{2}$ is satisfied. 
Equation (\ref{eq:width_Voigt_f_sum}) and the simplified expressions given by Eqs.~(\ref{eq:spectrum_s}) and (\ref{eq:spectrum_f}) contribute to the main results of this paper. 

In Fig. ~\ref{fig:corr_spec}(a), we compare the exact result obtained from Eq.~(\ref{eq:spincorr_e}) and the approximate result obtained from Eq.~(\ref{eq:spincorr_m}). 
The results are shown for $\Gamma_{s}=0.01$; for a fixed value of $\Gamma$, the approximation is better as $\Gamma_{s}$ increases. 
We also present the result for a short-time expansion given by Eq.~(\ref{eq:spincorr_f1}).
We find that the degree of the approximation improves by modifying Eq.~(\ref{eq:spincorr_f1}) as in Eq.~(\ref{eq:spincorr_m}). 
In Fig. ~\ref{fig:corr_spec}(b), we compare the numerically exact spectrum obtained using Eq.~(\ref{eq:spincorr_e}) with Eq.~(\ref{eq:spinspe}) and the approximate one obtained using Eq.~(\ref{eq:width_Voigt_f_sum}) by summing the first four terms. 
The results are shown again for $\Gamma_{s}=0.01$; the approximation is better as $\Gamma_{s}$ increases for the same value of $\Gamma$. 
When $\Gamma$ is varied, the convergence of the sum is obtained for $\Gamma>1/\sqrt{2}$ regardless of the values of $\Gamma_{s}$.

%%%
\section{Results of FWHM}
\label{sec:IV}
\begin{figure}[h]
\begin{center}
    \includegraphics[width=0.5\textwidth]{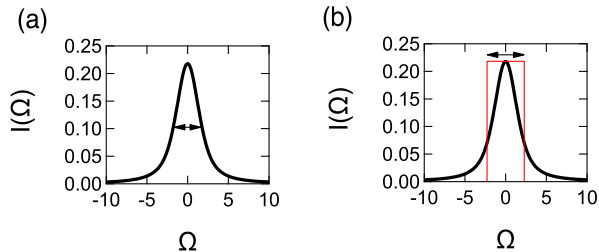}
\end{center}
\caption{(Color online) Frequency dependence of the power spectrum illustrating (a) FWHM and (b) the equivalent  width (integral breadth). FWHM is defined by the width when the height of the spectrum is half of its maximum value. The equivalent width is defined using the maximum height and the area under the spectrum as a function of the angular frequency. The kurtosis parameter is defined by the equivalent width divided by FWHM. The spectral lines are drawn for $\Gamma_{s}=1$ and $\Gamma=1$.}
\label{fig:FWHM_EW}
\end{figure}

We study the FWHM and the equivalent width as (or the integral breadth) indicated in Fig. ~\ref{fig:FWHM_EW}. 
The equivalent width can be found by drawing a rectangle with a height equal to that of the absorption line shape and the area equal to that under the absorption line; the width of the rectangle thus specified is called the equivalent width. 
When the area is normalized, the equivalent width is equal to the inverse of the height. 
The kurtosis parameter 
(the inverse of the form factor) of the spectrum, defined by the equivalent width divided by the FWHM, \cite{Langford_78} reflects the spectral shape variation by motional narrowing. 
The kurtosis parameter value is larger when the kurtosis of the peak is larger. 
The kurtosis parameter of the Lorentzian spectrum is $\pi/2\approx1.57$ and that of the Gaussian spectrum is $\sqrt{\pi/\log2}/2\approx1.06$. 
The kurtosis parameter value of the Lorentzian spectrum is larger than that of the Gaussian spectrum. 

Using the empirical equation given by $ r+\sqrt{r^2+8 \ln(2) \sigma^2}$ to express the FWHM of the Voigt function $V(x; \sigma,r)$,\cite{OLIVERO_77} the FWHM of the Voigt function given in Eq.~(\ref{eq:spectrum_s}) for small $\gamma$ is written
\begin{align}
f_{Vw} &\approx 2\left(\gamma_{s}+\sqrt{\gamma_{s}^2+2 \ln(2) \Delta_0^2} \right).
\label{eq:width_Voigt_s}
\end{align}
From Eq.~(\ref{eq:spectrum_f}), the FWHM of the Voigt function given in Eq.~(\ref{eq:spectrum_f}) for large $\gamma$ is narrower than the above result and is expressed as 
\begin{align}
f_{Vn} &\approx 2\left(\gamma_{s}+\sqrt{\gamma_{s}^2+\ln(2) \Delta_0^2} \right). 
\label{eq:width_Voigt_f}
\end{align}

\begin{figure}[h]
\begin{center}
    \includegraphics[width=0.5\textwidth]{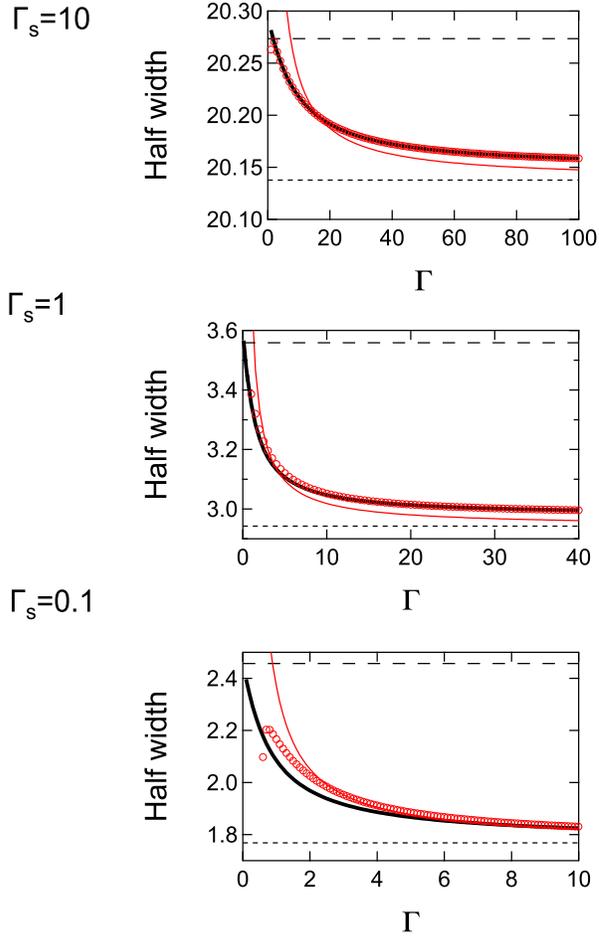}
\end{center}
\caption{(Color online) Normalized FWHM using $\Delta_0$ is shown against $\Gamma=\gamma/\Delta_0$. The thick solid lines represent the numerical results of Eq.~(\ref{eq:spinspe}) using Eq.~(\ref{eq:spincorr_m}). The long dashed lines and dots represent the results of Eq.~(\ref{eq:width_Voigt_s}) and those of Eq.~(\ref{eq:width_Voigt_f}), respectively. The (red) thin lines indicate $\Gamma$--dependence given by Eq.~(\ref{eq:narrowgd}). The (red) circles indicate FWHM calculated from the approximate expression given by Eq.~(\ref{eq:width_Voigt_f_sum}).
}
\label{fig:f}
\end{figure}
In Fig. ~\ref{fig:f}, we show the FWHM normalized by $\Delta_0$ as a function of the dimensionless transition rate. 
The exact numerical results were calculated by substituting Eq.~(\ref{eq:spincorr_m}) into Eq.~(\ref{eq:spinspe}) and performing a numerical integration. 
Judging from Fig. ~\ref{fig:f}, we note that the fraction of narrowing effect is larger when $\Gamma_{s}$ is smaller. 
When $I(\Omega)$ is dominated by the first term given by $j=0$ of Eq.~(\ref{eq:width_Voigt_f_sum}), the FWHM can be approximated by 
\begin{align}
f_{V} &\approx \gamma_{s}+\frac{\Delta_0^2}{2\gamma}+\sqrt{\left(\gamma_{s}
+\frac{\Delta_0^2}{2\gamma}\right)^2+4\ln(2) \Delta_0^2}.
\label{eq:narrowgd}
\end{align}
The (red) thin lines indicate the results of the approximate solution given by Eq.~(\ref{eq:narrowgd}). 
We find that the approximation is satisfactory for $\Gamma>\mbox{Max}(\Gamma_{s},1)$. 
The (red) circles represent the FWHM calculated numerically using Eq.~(\ref{eq:width_Voigt_f_sum}) by summing the first four terms. 
As long as $\Gamma>1/\sqrt{2}$ is satisfied, the red circles approximate the exact numerical results.
Because the integration of the spectrum over the angular frequency is normalized by $1$, the equivalent width can be obtained from $1/I(0)$. 
\begin{figure}[h]
\begin{center}
    \includegraphics[width=0.5\textwidth]{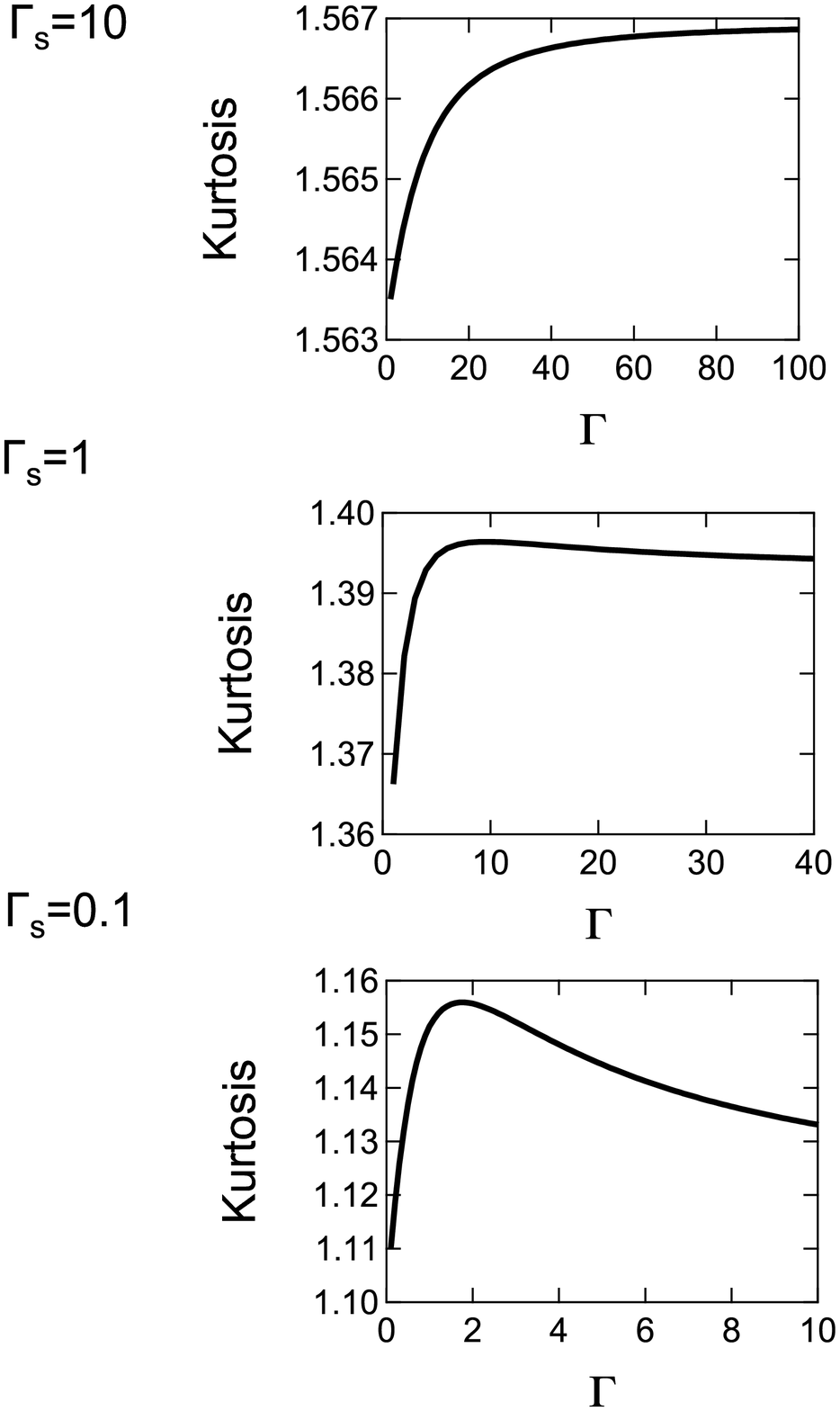}
\end{center}
\caption{The kurtosis parameter is shown as a function of $\Gamma$ for various values of $\Gamma_s$. The kurtosis parameter values are 1.06 and 1.57 for the Gaussian spectrum and the Lorentzian spectrum, respectively.}
\label{fig:Kurtosis}
\end{figure}
We define the kurtosis parameter by the equivalent width divided by the FWHM. \cite{Langford_78}
In Fig.  \ref{fig:Kurtosis}, the kurtosis parameter is shown against $\Gamma$ for various values of $\Gamma_s$.
The variation of the kurtosis parameter values against $\Gamma$ is larger as the value of $\Gamma_s$ is smaller. 
For $\Gamma_s=10$, we find a monotonic increase in the kurtosis parameter as $\Gamma$ increases; the line shape approaches a Lorentzian line shape through motional narrowing. 
For $\Gamma_s=0.1$, we find the maximum value of the kurtosis parameter as a function of $\Gamma$. 
The appearance of the maximum and the relatively small kurtosis parameter value at the maximum indicate that the narrowing is not the qualitative change in the spectrum from the Gaussian function to the Lorentzian function but the Gaussian-like shape is retained during the narrowing process. 

\begin{figure}[h]
\begin{center}
    \includegraphics[width=0.5\textwidth]{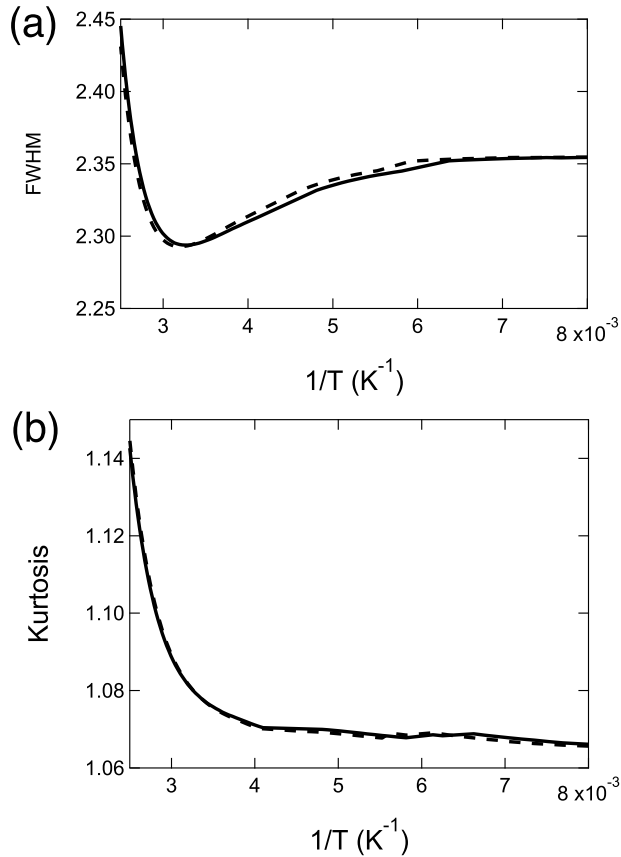}
\end{center}
\caption{(a) FWHM normalized by $\Delta_0$ and (b) kurtosis parameter plotted against inverse temperature. The kurtosis parameter values are $1.06$ and $1.57$ for the Gaussian spectrum  and the Lorentzian spectrum , respectively. The dashed lines are obtained for $E_a=0.075$~eV, and the solid lines indicate the results of the Marcus equation for $\lambda=0.3$ ~eV. The parameter values are $c_s=1025.86$, $c=1.79$, $E_{as}=0.3$~eV, and $T_m=300$~K for the Marcus equation.}
\label{fig:temp}
\end{figure}
In Fig. ~\ref{fig:temp}(a), we present the temperature variation of the spectral width. 
We assume that the on-site 
%%% revised 
inherent 
%%% revised 
 (nonadiabatic) relaxation is 
due to thermal activation with $\Gamma_{s}=c_{s} \exp[-E_{as}/(k_{\rm B}T)]$,\cite{Orbach_62} where $E_{as}$ denotes the activation energy, $k_{\rm B}$ the Boltzmann constant,  
$T$ the temperature, and $c_{s}$ a parameter. 
For a given hopping transition rate, we assume thermal detrapping from a trap given by $\Gamma=c \exp[-E_{a}/(k_{\rm B}T)]$\cite{Kehr_96} with $E_{a}$ the activation energy and $c$ a parameter. 
For another hopping transition rate, we assume the Marcus equation described by $\Gamma=c \sqrt{T_m/T}\exp[-\lambda/(4k_{\rm B}T)]$,\cite{Marcus_56,Marcus_64,Levich_59, Holstein_59, HOLSTEIN_59_2} with $\lambda$ the reorganization energy and $T_m$ a parameter. 
In regard to the temperature dependence, the difference between these two transition rates turns out to be small (see Fig. ~\ref{fig:temp}). 
As the temperature is increased from a low temperature limit, the minimum of the width appears as a result of motional narrowing. 
In the high temperature regime, the increase in width obtained by increasing the temperature originates from the temperature dependence of the on-site relaxation rate constant denoted by $\Gamma_{s}$. 
The similar minimum of the FWHM as a function of inverse temperature is observed experimentally in organic 
%%% revised
materials.\cite{Matsui_08,Matsui_12,Matsui_10}
%%% revised 
In Fig. ~\ref{fig:temp}(b), we show the corresponding values of the kurtosis parameter as a function of inverse temperature. 
Judging from these values, the line shape is closer to a Gaussian profile than a Lorentzian profile 
over the entire temperature range. 
In the above analysis, the on-site  
%%% revised 
inherent 
%%% revised
(nonadiabatic) relaxation rate 
was kept small within the temperature variation and 
this is the reason that the near-Gaussian profile is retained. 
If the on-site 
%%% revised 
inherent 
%%% revised 
 (nonadiabatic) relaxation rate is large, as characterized by $\Gamma_{s}\gg1$, 
the narrowing of a near-Lorentzian line shape can be obtained judging from Figs. \ref{fig:f}-\ref{fig:Kurtosis}. 
For such parameters, the profile is closer to being Lorentzian but the variation of the FWHM becomes smaller, as indicated in Fig. ~\ref{fig:f}. 

%%%
\section{First derivative of the power spectrum}
\label{sec:V}
\begin{figure}[h]
\begin{center}
    \includegraphics[width=0.5\textwidth]{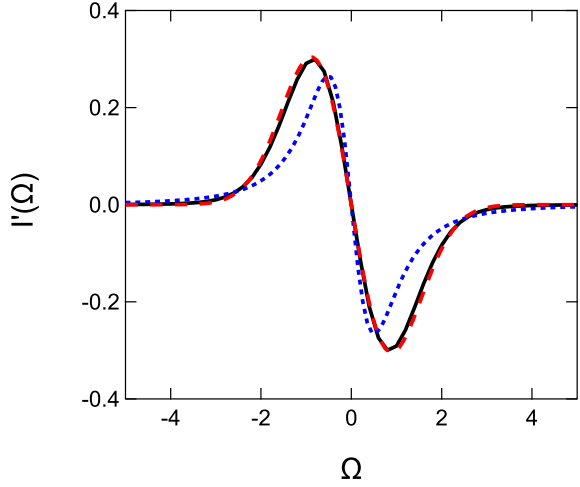}
\end{center}
\caption{(Color online) First derivative of the absorption spectrum plotted against angular frequency for $\Gamma_{s}=0.01$ and $\Gamma=1$. The solid line represents the convergent values of Eq.~(\ref{eq:width_Voigt_f_suml}). The summation of the first four terms yields the convergent results. The (red) dashed and (blue) dots indicate the fitting of the solid line by a Gaussian function and a Lorentzian function, respectively. (We find $\sigma=0.891$ for the Gaussian function [Equation~(\ref{eq:G})] and $r=0.882$ for the Lorentzian function [Equation~(\ref{eq:L})].) The kurtosis parameter value is $1.11$.}
\label{fig:firstderivatives}
\end{figure}

When the recorded ESR spectra are the first derivatives of the absorption spectra, it is more convenient to calculate the first derivative of the spectrum to compare directly with the measured spectrum. 
This first derivative is calculated using the derivative of the Voigt function given by\cite{NIST}
\begin{align}
\frac{\partial}{\partial x} V(x; \sigma,r)&=- \mbox{Re}\left[
\frac{x+i r}{\sqrt{2\pi}\, \sigma^3} \exp\left(-\frac{(x-ir)^2}{2\sigma^2} \right)
\mbox{erfc} \left(\frac{(ix+r)}{\sqrt{2}\, \sigma}\right) \right]
\label{eq:spectrum_sd}\\
&=\frac{-x V(x; \sigma,r)+ rL(x; \sigma,r)}{\sigma^2}, 
\label{eq:spectrum_sd1}
\end{align}
where the imaginary Voigt distribution function is given by,\cite{NIST}
\begin{align}
L(x; \sigma,r)=\mbox{Im}\left[\exp\left(-\frac{(x-ir)^2}{2\sigma^2} \right)
\mbox{erfc} \left(\frac{(ix+r)}{\sqrt{2}\sigma}\right) \right]/\sqrt{2\pi \sigma^2}.
\label{eq:Voigt_L}
\end{align}
Using Eq.~(\ref{eq:width_Voigt_f_sum}), the first derivative of the power spectrum is obtained 
\begin{align}
I^{\prime}(\Omega) &=\sum_{j=0}^\infty \frac{1}{j!} \left(-\frac{1}{2\Gamma^2} \right)^j 
\exp\left(\frac{1}{2\Gamma^2} \right) \nonumber \\
&\frac{-\Omega V\left(\Omega; \frac{1}{\sqrt{2}},\Gamma_{s}
+\frac{1}{2\Gamma}+\Gamma j \right)+(\Gamma_{s}+\frac{1}{2\Gamma}+\Gamma j)
L\left(\Omega; \frac{1}{\sqrt{2}},\Gamma_{s}+\frac{1}{2\Gamma}+\Gamma j \right)}{\sigma^2}. 
\label{eq:width_Voigt_f_suml}
\end{align}
In the limit of small and large values of $\gamma$, the first derivative of the power spectrum is given by 
\begin{align}
I^{\prime}(\omega)&\approx \frac{-\omega V(\omega; \Delta_0,2\gamma_{s})+ 2\gamma_{s}L(\omega; \Delta_0,2\gamma_{s})}{\sigma^2}, \\
I^{\prime}(\omega)& \approx \frac{-\omega V(\omega; \Delta_0/\sqrt{2},2\gamma_{s})+ 2\gamma_{s}L(\omega; \Delta_0/\sqrt{2},2\gamma_{s})}{\sigma^2},
\label{eq:spectrum_sl}
\end{align}
respectively, where we have used Eqs.~(\ref{eq:spectrum_s}) and (\ref{eq:spectrum_f}). 
In Fig. ~\ref{fig:firstderivatives}, we plotted the first derivative of the power spectrum when the Gaussian spectrum is partially narrowed. 
To avoid the effect of on-site nonadiabatic relaxation rate constant given by $\Gamma_{s}$, we set a small value for $\Gamma_{s}$. 
The value of $\Gamma$ is chosen so that partial narrowing of the Gaussian spectrum occurs (see Fig. ~\ref{fig:f}). 
We find that the first derivative of the power spectrum calculated from Eq.~(\ref{eq:width_Voigt_f_suml}) is close to the first derivative of a Gaussian profile compared with that of a Lorentzian profile. 
The result is consistent with the low kurtosis parameter value given by $1.11$. 

\section{Generalization to disordered systems}
\label{sec:VI}
%%%%%%%%%%%%%%%%%%%%%%%%%%%%%%%%%%%%%%
So far, we have considered Markovian transitions. 
Non-Markovian transitions have been considered for the Kubo-Anderson model using waiting time distribution functions. 
\cite{Jung_02,Loring_85,Sanda_05} 
Here, we take into account 
the averaging with respect to local field at each site in addition to 
Non-Markovian transitions.  

In the Markovian model, 
we have considered the case that the transitions between the sites are characterized by a rate constant associated with 
a single activation energy. 
When the activation energy is distributed, 
the waiting time distribution for the transitions becomes non-exponential. 
We consider the waiting time distribution of spin at site a for the transition to site b denoted by $\psi (t)$. 
For simplicity, we consider the same waiting time distribution of spin at site b. 
Algebraic asymptotic tails could result from the distribution of the activation energy. \cite{Scher,Jakobs_93}
When the activation energy distribution function  
is the exponential function with the energy depth given by $k_{\rm B} T_{\rm c}$, \cite{Scher} 
$
g(E)= \exp \left[ - E/(k_{\rm B} T_{\rm c}) \right] /(k_{\rm B} T_{\rm c}) 
$, and the activated release rate is assumed for transition between the sites, 
$
\gamma (E) = \gamma_{\rm r} \exp \left[ - E/(k_{\rm B} T) \right] 
$, 
the waiting time distribution is expressed as  \cite{Scher}
\begin{eqnarray}
\psi (t) = \int_0^{\infty} d\,E g(E) \gamma (E) 
%e^{- \gamma (E) t}
\exp \left( - \gamma (E) t \right)
\sim \frac{\alpha \Gamma \left( \alpha + 1\right)}{\gamma_{\rm r}^\alpha t^{\alpha+1}} , 
\end{eqnarray}
where $\alpha \equiv T/T_{\rm c}$ and $\Gamma (z)$ is the Gamma function. \cite{NIST}
The waiting time distribution shows the power law decay with the absolute value of the exponent given by $\alpha+1$. 
Owing to the activation energy distribution, the waiting time distribution exhibits heavy tail property.  
By decreasing the temperature, the absolute value of the exponent becomes lower. 
The effect of heavy tail property on the diffusion has been studied by the continuous time random walk model using the waiting time distribution function for transition to a neighboring site. \cite{Montroll}
According to the continuous time random walk model, 
the exponent $\alpha$ characterizes the subdiffusive growth of the mean square displacements given by 
$t^\alpha$, where $\alpha=1$ for the normal diffusion and $\alpha <1$ for the dispersive diffusion. 
%We denote the Laplace transform of an arbitrary function $f(t)$ by $\hat{f} (s)$. 
In the small $s$ limit, the Laplace transform 
of the waiting time distribution function is given by \cite{Scher}
\begin{eqnarray}
\hat{\psi} (s) 
\sim 1 - 
\left[ \pi \alpha/\sin (\pi \alpha) \right] \left( s/ \gamma_{\rm r} \right)^{\alpha} 
\mbox{ for }  s/\gamma_{\rm r} < 1 .
\label{eq:dispersive1}
\end{eqnarray}
The asymptotic expression will be used below. 
In this section, we consider the waiting time distribution function having an algebraic asymptotic tail and being normalized; 
 $\int_0^\infty d\, t\psi(t)=1$. 

As before, 
we calculate the spin correlation averaged over the distribution of the local Larmor angular frequency and the initial condition,  
where the even site occupation probability is assumed.  
%%% revised 
As shown in the Appendix B, 
%%% revised  
the spin correlation function becomes
\begin{align}
\phi(t) =\exp \left( - \frac{1}{4}\Delta_0^2 t^2 - 2 \gamma_{s} t \right) 
\left[1+
\sum_{m=1}^\infty \frac{(-1)^m}{m!} \left(\frac{\Delta_0 t}{2} \right)^{2m} 
\frac{(2m)!}{(m-1)!} E_{\alpha,2m+1-\alpha(m-1)}^{(m-1)}
\left(-\gamma_\alpha t^\alpha \right)
\right].
\label{eq:disorder15}
\end{align}
This is the exact solution and is generalization of Eq.~(\ref{eq:spincorr_e}) when the waiting time distribution has heavy tail property due to, 
{\it e.g.}, the distribution of the activation energy.  
Equation (\ref{eq:disorder15}) is the main result of this section.

For a slow transition rate expressed by $\gamma_\alpha t^\alpha \ll 1$, 
Eq.~(\ref{eq:disorder15}) can be simplified by introducing the lowest order series expansion given by 
$E_{\alpha,3} \left(-\gamma_\alpha t^\alpha \right) \approx 1/2$ as
\begin{align}
\phi_{s}(t) \approx \exp \left( - \frac{1}{2}\Delta_0^2 t^2 - 2 \gamma_{s} t \right). 
\label{eq:disorder16}
\end{align}
In contrast, for a fast transition rate expressed by $\gamma_\alpha t^\alpha \gg 1$, taking the limit reduces Eq.~(\ref{eq:disorder15}) to 
\begin{align}
\phi_{f}(t) \approx \exp \left( - \frac{1}{4}\Delta_0^2 t^2 - 2 \gamma_{s} t \right). 
\label{eq:disorder17}
\end{align}
As explained below Eq.~(\ref{eq:spincorr_f}), 
this equation indicates a slowing down of the spin relaxation and narrowing of the absorption spectral line. 
It should be noted that the condition $\gamma_\alpha t^\alpha \gg 1$ could be hardly attained when the value of $\alpha$ is lowered; 
the results suggest that the motional narrowing can be hardly observed when the hopping motion is sluggish in disordered media. 

A correction to Eq.~(\ref{eq:disorder17}) may be calculated from Eq.~(\ref{eq:disorder15}), the result being
\begin{align}
\phi_f (t) \approx \exp \left( - \frac{1}{4}\Delta_0^2 t^2 - 2 \gamma_{s} t \right)
\left[1-\frac{(\Delta_0 t)^2}{2} E_{\alpha,3}
\left(-\gamma_\alpha t^\alpha \right) \right], 
\label{eq:disorder18}
\end{align}
where the summation in Eq.~(\ref{eq:disorder15}) is taken up to $m=1$. 
As before Eq.~(\ref{eq:disorder18}) can be modified to 
\begin{align}
\phi (t) \approx \exp \left[ - \frac{1}{4}\Delta_0^2 t^2
- 2 \gamma_{s} t -\frac{(\Delta_0 t)^2}{2} E_{\alpha,3}
\left(-\gamma_\alpha t^\alpha \right) \right]. 
\label{eq:disorder20}
\end{align}
By setting $\alpha=1$ and using  
$ E_{1,3}
\left(-x \right)=[x-1+\exp(-x)]/x^2$, 
Eq.~(\ref{eq:disorder20}) reduces to Eq.~(\ref{eq:spincorr_m}),  
where $\gamma=\gamma_1$. 
By applying the asymptotic expansion expressed by 
\begin{align}
E_{\alpha,3}
\left(-\gamma_\alpha t^\alpha \right) \approx \frac{1}{\Gamma(3-\alpha)\gamma_\alpha t^\alpha} ,
\label{eq:Easym}
\end{align}
the asymptotic expression of Eq.~(\ref{eq:disorder20}) can be given by 
 \begin{align}
\phi (t) \approx \exp \left[ - \frac{1}{4}\Delta_0^2 t^2
- 2 \gamma_{s} t -\frac{\Delta_0^2 t^{2-\alpha}}{2\Gamma(3-\alpha)\gamma_\alpha } \right]. 
\label{eq:disorder20_a}
\end{align}
The last term in Eq.~(\ref{eq:disorder20_a}) indicates the exponential decay only when $\alpha=1$. 
For $\alpha<1$, non-exponential decay can be obtained 
owing to heavy tail property in the waiting time distribution. 
The exponent approaches to $2$ as $\alpha$ decreases; 
the exponential decay for $\alpha=1$ changes to the Gaussian-like decay as $\alpha$ decreases.  
Therefore, the corresponding power spectrum will deviate from Lorentzian form when $\alpha<1$.

In Eq.~(\ref{eq:disorder20}) 
%%% revised 
[See also Eqs. (\ref{eq:delta}) and (\ref{eq:2018_2_6_2_2})], 
%%% revised 
$\langle \phi_r(t)\rangle_\delta$ is approximated by 
\begin{align}
\langle \phi_r(t)\rangle_\delta \approx \exp \left[ -\frac{(\Delta_0 t)^2}{2} E_{\alpha,3}
\left(-\gamma_\alpha t^\alpha \right) \right]. 
\label{eq:phirappr}
\end{align}
It can be rewritten as 
\begin{align} 
\langle \phi_r(t)\rangle_\delta \approx \exp  \left[ -\Delta_0^2 \int_0^t d t_1\, 
(t-t_1) \psi(t_1)
\right], 
\label{eq:phirappr1}
\end{align}
where $\psi(t)$ corresponds to the correlation function when the fluctuating angular frequency is modeled as 
the Gaussian-Markovian fluctuation. 
$\psi(t)$ is given by 
\begin{align} 
\psi(t)=E_{\alpha}
\left(-\gamma_\alpha t^\alpha \right) , 
\label{eq:psi1}
\end{align}
where the Mittag-Leffler function $E_{\alpha}(-\gamma_\alpha t^\alpha)=E_{\alpha,1}(-\gamma_\alpha t^\alpha)$ is 
defined through the inverse Laplace transformation as 
$E_{\alpha}(-\gamma_\alpha t^\alpha)={\cal L}^{-1} \left[ 1/\left(s+\gamma_\alpha s^{1-\alpha} \right) \right]$.
Initial decay obeys the stretched exponential form 
$ 
\psi(t)=\exp
\left[-\gamma_\alpha t^\alpha/\Gamma(1+\alpha)  \right] 
$, 
which suggests that the Mittag-Leffler function is the natural generalization of the exponential 
correlation function. 
Indeed, $E_{\alpha}(-\gamma_\alpha t^\alpha)$ represents a solution of the fractional relaxation equation given by
$(\partial/\partial t)\psi(t)= - \gamma_\alpha \mbox{ }_0 D_t^{1-\alpha} \psi(t)
$, \cite{METZLER_01}
where the Riemann-Liouville fractional derivative is defined through the convolution to an arbitrary function $f(t)$ as 
$
\mbox{ }_0 D_t^{1-\alpha}f(t)=[1/\Gamma ( \alpha )](\partial/\partial t)
\int_0^t d\,t_1 f(t_1) /\left( t - t_1 \right)^{1-\alpha} 
$. 
The conventional exponential relaxation function is obtained for $\alpha=1$. 
On the other hand, the asymptotic expansion given by 
$\psi(t)=\left(\Gamma(1-\alpha)\gamma_\alpha t^\alpha\right)^{-1}
$ leads to 
$\langle \phi_r(t)\rangle_\delta \approx \exp \left[ -(\Delta_0^2 t^{2-\alpha})/\left\{2\Gamma(3-\alpha)\gamma_\alpha \right\} \right]$, 
which shows intrinsic non-exponential decay when $\alpha<1$. 

The conclusion of this section is as follows. 
In disordered media, power-law waiting time distribution is obtained when the activation barrier for hopping is exponentially distributed. 
The initial decay of the spin correlation function was Gaussian function of time 
and was not influenced by the presence of heavy tail property in the waiting time distribution of 
hopping transitions. 
The asymptotic decay was influenced by the presence of heavy tail property in the waiting time distribution;  
by increasing heavy tail property in the waiting time distribution function, 
the exponential-like decay component influenced by the hopping frequency 
approaches to the Gaussian-like decay.  
The results indicate that the motional narrowing to Lorentzian absorption line shape could be hardly observed 
in disordered media, though the narrowing of Gaussian absorption line shape 
while retaining the Gaussian-like line shape could be possible. 
Moreover, the on-set time of slowing down of spin relaxation increases with decreasing 
the absolute value of the exponent of the power-law waiting time distribution; 
motional narrowing phenomena could be hardly observed when $\alpha \ll 1$.  

%%%
\section{Conclusion}
\label{sec:VII}

We studied motional narrowing of spins using the Kubo--Anderson--Kitahara model. 
In the model, the transition between two sites and the distribution of local field on each site were explicitly taken into account. 
The local field distribution of spins was assumed to be Gaussian. 
The transition between the sites having an inhomogeneous local field induces an adiabatic relaxation of the spin. 
We also considered the on-site 
%%% revised 
inherent 
%%% revised
 (nonadiabatic) relaxation of the spin. 

We obtained the exact solution of the spin correlation function. [Equation~(\ref{eq:spincorr_e})]
We also derived an approximate expression given by Eq.~(\ref{eq:spincorr_m}); 
the spin correlation function was expressed by a well-known function describing the narrowing of a Gaussian profile with a multiplicative correction term arising from the transition between sites under inhomogeneous local field environments.\cite{Kubo_Tomita,Kubo_54}
Because the spin correlation function was expressed in the form similar to the well-known function derived by assuming a Gaussian--Markovian fluctuation of the field, the narrowing process caused by transitions between sites having an inhomogeneous local field can be understood based on Gaussian--Markovian field fluctuations. 
More importantly, some deviations from this simplified view are also clearly recognized. 
The additional term $\exp\left[-(1/4)\Delta_0^2 t^2\right]$ can be regarded as an ensemble average of the mean angular frequency between the sites. 
The effect of carrier transitions between the sites having distributed angular frequency gives rise to inhomogeneous broadening in the ESR spectrum. 
When the distribution of the on-site local field is Gaussian, the relaxation is given by a Gaussian function. 
%%% revised 
As we discussed below Eq. (\ref{eq:spincorr_f}), 
recursive nature of transitions manifest itself in the spin correlation function in the fast-transition-rate limit. 
For extended systems, recursive nature of transitions should depend on the dimensionality and the lattice structure. 
Unfortunately,  the exact analytical treatment is exceedingly difficult for spatially extended systems.
%%% revised

The motional narrowing has been studied using the absorption spectrum rather than the spin correlation function. 
We derived the approximate expression of the absorption spectrum in terms of the Voigt function. 
The Voigt function, as a Gaussian--Lorentzian convolution, is suited to describing the combined effect of Gaussian line-shape and the Lorentzian line-shape broadening. 
Using an empirical equation describing FWHM of the Voigt function,\cite{OLIVERO_77} we examined the motional narrowing for 
a wide range of values of the on-site 
%%% revised 
inherent 
%%% revised 
 (nonadiabatic) relaxation rate constant. 
We also analyzed the kurtosis parameter defined using the equivalent width and the FWHM. 
The value of the kurtosis parameter increases as the Gaussian line shape changes to a Lorentzian line shape. 

When the on-site 
%%% revised 
inherent 
%%% revised 
 (nonadiabatic) relaxation rate is small, the Gaussian line shape changes to a narrowed Gaussian line shape by increasing the transition rate. 
The result is confirmed by studying the first derivative of the absorption spectrum. 
The kurtosis parameter shows a maximum deviation from the Gaussian line shape in the course of narrowing. 

When the on-site 
%%% revised 
inherent 
%%% revised 
(nonadiabatic) relaxation rate is large, the kurtosis parameter increases monotonically by increasing the transition rate; 
the line shape approaches a Lorentzian line shape through motional narrowing.

The fraction of the FWHM variation plotted against its maximum value decreases by increasing the on-site 
%%% revised 
inherent  
%%% revised 
(nonadiabatic) relaxation rate. 
Therefore, the narrowing effect is larger by decreasing the on-site 
%%% revised 
inherent 
%%% revised 
 (nonadiabatic) relaxation rate. 
By assuming activation type for both adiabatic and nonadiabatic relaxation rates, the FWHM shows a minimum as a function of inverse temperature; the result is consistent with the experimental results.\cite{Matsui_08,Matsui_12,Matsui_10}

We also considered 
non-Markovian transitions using power-law waiting time distribution. 
In disordered media, 
power-law waiting time distribution results from the exponential distribution of activation barrier for hopping. 
The initial decay of the spin correlation function was not 
influenced by the presence of heavy tail property in the waiting time distribution. 
The asymptotic decay component influenced by the hopping frequency 
approaches to Gaussian-like decay by increasing heavy tail property in the waiting time distribution function.  
Moreover, the on-set time of slowing down of spin relaxation increases 
as the hopping motion is sluggish in disordered media. 

For simplicity, we only considered two-site transitions. 
As pointed out previously, there may be an extra subtle effect for extended systems and the results would depend on the dimensionality.\cite{Czech_84,Czech_86,Mitra_91}
When the mobility of a spin carrier is high, such effects may be important. 
We also assumed the Gaussian random local field to be in the direction of the external field. 
Although we considered a simplified model, our results captured an additional effect often ignored in assuming Gaussian--Markovian field fluctuations by explicitly considering adiabatic transitions in locally inhomogeneous field environments. 

The motional narrowing effect has been observed in the optical resonance absorption line shapes as well as the magnetic 
resonance absorption line shapes. \cite{Sumi_77,Blumen_78,Jackson_81,Cao_09,Rajesh_17}
In optical line shapes, the quantum coherent effects on line shapes give additional difficulties to analyze the motional 
narrowing effect. 
The role of quantum coherence in the narrowing of optical line shape under both the site energy fluctuation and the off-diagonal coupling between sites has been treated recently. \cite{Rajesh_17}
As a future problem, it is necessary to take quantum coherence into account to study the narrowing effect on the optical line shapes.  

%%%
\acknowledgments

This work was supported by JSPS KAKENHI Grant Number 15K05406. 
We would like to thank Professor K. Kitahara for informing us Ref. \onlinecite{KOK_87}.

\newpage
\renewcommand{\theequation}{A.\arabic{equation}}  
\setcounter{equation}{0}  % reset counter  
\section*{Appendix A. Derivation of Eq.~(\ref{eq:spincorr_e})}
%\section*{Appendix A. }
We define $M(t)=M_a(t)+M_b(t)$ and $q(t)=M_a(t)-M_b(t)$. 
The equations of motion for $M(t)$ and $q(t)$ are obtained as 
\begin{align}
\frac{d}{dt} M(t)&=i \omega_m M(t) - \gamma_{s} M(t)+ i \delta q (t)
\label{eq:M1_1}\\
\frac{d}{dt} q(t)&=i \omega_m q(t) - \gamma_{s} q(t)+i \delta M(t) -\gamma q(t),
\label{eq:M1_2}
\end{align}
where the initial condition is given by $M(0)=M_0$ and $q(0)=0$; $\omega_m$ and $\delta$ are defined by 
\begin{align}
\omega_m=\frac{1}{2} \left(\omega_a+\omega_b\right),
\label{eq:omegam}\\
\delta=\frac{1}{2} \left(\omega_a-\omega_b\right).
\label{eq:delta}
\end{align}

Using the Laplace transformation, defined by $\hat{f}(s)=\int_0^\infty dt \exp(-st) f(t)$ for a function $f(t)$, and eliminating $\hat{q}(s)$ by taking the Laplace transformation of Eqs.~(\ref{eq:M1_1})--(\ref{eq:M1_2}), we obtain 
$\hat{\phi}(s)=\hat{\phi}_r(s-i \omega_m+ 2 \gamma_s)$, where the Laplace transform of $\phi_r (t)$ is given by 
\begin{align}
\hat{\phi}_r(s)=\frac{s+\gamma}{s(s+\gamma)+\delta^2} .
\label{eq:2018_2_6_2_5}
\end{align}
By the inverse Laplace transformation, we obtain
\begin{align}
\phi(t)=\exp\left[\left(i \omega_m-2\gamma_s \right)t\right]
\phi_r (t). 
\label{eq:M_4}
\end{align}

The term $\langle \cdots \rangle$ in Eq.~(\ref{eq:spincorr}) can be calculated by first averaging with respect to $\omega_m$ and subsequently averaging with respect to $\delta$; the averaging procedure can be expressed as $\langle \langle \cdots \rangle_{\omega_m} \rangle_\delta$, where each average is given by 
\begin{align}
\langle \cdots \rangle_{\omega_m} =\frac{1}{\sqrt{\pi}\, \Delta_0}
\int_{-\infty}^\infty d \omega_m \exp\left(- \frac{\omega_m^2}{\Delta_0^2} \right) \cdots
\label{eq:2018_2_6_2_1}\\
\langle \cdots \rangle_\delta = \frac{1}{\sqrt{\pi}\, \Delta_0}
\int_{-\infty}^\infty d \delta \exp \left(- \frac{\delta^2}{\Delta_0^2} \right) \cdots.
\label{eq:2018_2_6_2_2}
\end{align}

The averaging of $\phi(t)$ with respect to $\omega_m$ is obtained as 
\begin{align}
\langle \phi(t)\rangle_{\omega_m}
=\exp\left(-\frac{1}{4} \Delta_0^2 t^2 -2 \gamma_s t \right) \phi_r (t). 
\label{eq:2018_2_6_2_3}
\end{align}
The averaging with respect to $\delta$ can be calculated after the Laplace transformation. 
$\langle\hat{\phi}_r(s)\rangle_\delta$ is obtained as 
\begin{align}
\langle\hat{\phi}_r(s)\rangle_\delta=\sqrt{\frac{\pi}{\Delta_0^2}}\sqrt{\frac{s+\gamma}{s}}
\exp \left( \frac{s(s+\gamma)}{\Delta_0^2} \right) \mbox{erfc} \left(\sqrt{\frac{s(s+\gamma)}{\Delta_0}}\right)
\label{eq:M_14}
\end{align}
for $s>0$ and $\gamma>0$. 
Introducing the asymptotic expansion, 
\begin{align}
\mbox{erfc} z =\frac{\exp(-z^2)}{\sqrt{\pi}\, z}
\sum_{m=0}^\infty (-1)^m \frac{(2m)!}{m!(2z)^{2m}}
\label{eq:M17}
\end{align}
and denoting the inverse Laplace transformation by ${\cal L}^{-1}\left[\cdots \right]$, we then obtain 
\begin{align}
\langle \phi_r(t)\rangle_\delta
&={\cal L}^{-1} \left[\sum_{m=0}^\infty \left(-\frac{\Delta_0^2}{2}\right)^m 
\frac{(2m)!}{2^m m!}\, \frac{1}{s^{m+1}(s+\gamma)^m}\right] \\
&=\sum_{m=0}^\infty \frac{(-1)^m}{m!} \left(\frac{\Delta_0 t}{2} \right)^{2m} 
{\cal M} \left(m; 2m+1; -\gamma t \right),
\label{eq:2018_2_6_1}
\end{align}
where ${\cal M} \left(a; b; z\right)$ is the Kummer confluent hypergeometric function. \cite{NIST}
Finally, we obtain Eq. (\ref{eq:spincorr_e}).

\section*{Appendix B. Derivation of Eq.~(\ref{eq:disorder15})}
%\section*{Appendix A. }
Spin precession and relaxation during the waiting time denoted by $\Delta t$ can be expressed as 
$\exp\left[( i \omega_b- 2 \gamma_s) \Delta t \right]$ for the spin at site b. 
The occupation probability of spins at each site at time $t$ can be obtained by the time convolution of 
the time distribution of spin just arriving at the site and the remaining probability of the arrived spin. 
First, we note that  
the spin state $m_a(t)$ for the spin just arriving at site a at time t obeys the relation given by,  
\begin{align}
m_a (t)= \int_0^t d\, t_1 \exp\left[ (i \omega_b -2 \gamma_2)\left(t -t_1\right) \right] \psi\left(t-t_1\right) m_b(t_1) + \delta(t) m_a(0). 
\label{eq:disorder1}
\end{align}
Then, we calculate the remaining probability from the waiting time distribution as 
$\varphi(t)=1-\int_0^t d\, t_1 \psi(t_1)=\int_t^\infty d\, t_1 \psi(t_1)$; 
the remaining probability is the fraction of spin escaping from transition to site b until time $t$.  
The spin state at site a at time t is obtained as  
$M_a (t)= \int_0^t d\, t_1 \exp\left[( i \omega_a -2\gamma_s)\left(t -t_1\right) \right]\varphi (t-t_1) m_a(t_1)$, 
where the spin precession and relaxation during the remaining period are taken into account. 
We also define the similar properties for the site b. 

By applying Laplace transformation to Eq.~(\ref{eq:disorder1}) and subtracting 
$\hat{\psi}\left(s+i\omega_a+2\gamma_s\right) \hat{m}_a(s)$, we obtain, 
\begin{align}
\left[1- \hat{\psi}\left(s+i\omega_a+2\gamma_s\right)\right]\hat{m}_a (s)= \hat{\psi}\left(s+i\omega_b+2\gamma_s\right) \hat{m}_b(s) -\hat{\psi}\left(s+i\omega_a+2\gamma_s\right) \hat{m}_a(s)+ m_a(0). 
\label{eq:disorder2}
\end{align}
Using $\hat{M}_i(s)=\left[1- \hat{\psi}\left(s+i\omega_i+2\gamma_s\right)\right]\hat{m}_i (s)/\left(s+i\omega_i+2\gamma_s\right)$ for $i=a,b$, 
Eq.~(\ref{eq:disorder1}) can be rewritten as  
\begin{align}
\left(s+i\omega_i+2\gamma_s\right)\hat{M}_a (s)-m_a(0)= \frac{s\hat{\psi}\left(s+i\omega_i+2\gamma_s\right)}{1- \hat{\psi}\left(s+i\omega_i+2\gamma_s\right)} \hat{M}_b(s) -\frac{s\hat{\psi}\left(s+i\omega_i+2\gamma_s\right)}{1- \hat{\psi}\left(s+i\omega_i+2\gamma_s\right)} \hat{M}_a(s) . 
\label{eq:disorder3}
\end{align}
By introducing Eq.~(\ref{eq:dispersive1}), the inverse Laplace transform of Eq.~(\ref{eq:disorder3}) is obtained as  
\begin{align}
\frac{\partial}{\partial t} M_a (t)=& i\omega_a M_a(t)-2\gamma_s M_a(t)-\frac{\partial}{\partial t}
\int_0^t d\,t_1\frac{\gamma_\alpha}{2\Gamma ( \alpha )} \frac{\exp\left\{ (i \omega_a-2\gamma_s) \left(t -t_1\right) \right\}}{\left( t - t_1 \right)^{1-\alpha}} M_a (t_1) +
\nonumber \\
& \frac{\partial}{\partial t}
\int_0^t d\,t_1 \frac{\gamma_\alpha}{2\Gamma ( \alpha )} \frac{\exp\left\{( i \omega_b-2\gamma_s)\left(t -t_1\right) \right\}}{\left( t - t_1 \right)^{1-\alpha}} 
M_b (t_1) ,
\label{eq:disorder4}
\end{align}
where the generalized hopping frequency is defined by Eq.~(\ref{eq:disorder5}). 
By exchanging a and b we also obtain the similar equation for $M_b(t)$. 
Equation (\ref{eq:disorder5}) is the generalization of Eq.~(\ref{eq:spina}) when the waiting time distribution for transitions has heavy tail property. 
We assume the even site occupation probability of spins in equilibrium.
The equilibrium initial condition can be expressed as $M_a(0)=M_b(0)=M_0/2$.
As before, we introduce $M(t)=M_a(t)+M_b(t)$, $q(t)=M_a(t)-M_b(t)$, $\omega_m$ and $\delta$ defined by 
Eqs.~(\ref{eq:omegam})-(\ref{eq:delta}).
We calculate the spin correlation averaged over the distribution of the local Larmor angular frequency and the initial condition.  

%%% revised 
Using the Laplace transformation, we obtain $\hat{\phi}(s)=\hat{\phi}_r(s-i \omega_m+2\gamma_s)$ and 
\begin{align}
\hat{\phi}_r(s)=\frac{s+\gamma_\alpha s^{1-\alpha} }{
s(s+\gamma_\alpha s^{1-\alpha} )+\delta^2}, 
\label{eq:Laplacephir}
\end{align}
where the generalized hopping frequency is defined by 
\begin{align}
\gamma_\alpha\equiv \frac{\sin \pi \alpha}{\pi \alpha} \gamma_{\rm r}^{\alpha}.
\label{eq:disorder5}
\end{align}
By the inverse Laplace transformation, we obtain
\begin{align}
\phi(t)=\exp\left[\left(i \omega_m-2\gamma_s \right)t\right]
\phi_r (t) .
\label{eq:disorder6}
\end{align}

The averaging of $\phi(t)$ with respect to $\omega_m$ is obtained as 
\begin{align}
\langle \phi(t)\rangle_{\omega_m}
=\exp\left(-\frac{1}{4} \Delta_0^2 t^2 -2 \gamma_s t \right) \phi_r (t).
\label{eq:disorder9}
\end{align}
The averaging with respect to $\delta$ can be calculated after the Laplace transformation applying to  
$\phi_r (t)$. 
$\langle\hat{\phi}_r(s)\rangle_\delta$ is obtained from Eq.~(\ref{eq:Laplacephir}) as 
\begin{align}
\langle\hat{\phi}_r(s)\rangle_\delta=\sqrt{\frac{\pi}{\Delta_0^2}}\sqrt{\frac{s+\gamma_\alpha s^{1-\alpha}}{s}}
\exp \left( \frac{s(s+\gamma_\alpha s^{1-\alpha})}{\Delta_0^2} \right) \mbox{erfc} \left(\sqrt{\frac{s(s+\gamma_\alpha s^{1-\alpha})}{\Delta_0}}\right)
\label{eq:disorder12}
\end{align}
for $s>0$ and $\gamma_r>0$. 
By introduce the asymptotic expansion given by Eq.~(\ref{eq:M17}), we obtain 
\begin{align}
\langle \phi_r(t)\rangle_\delta
&=\sum_{m=0}^\infty \left(-\frac{\Delta_0^2}{2}\right)^m 
\frac{(2m)!}{2^m m!}\, {\cal L}^{-1} \left[ \frac{1}{s^{(2-\alpha)m+1}(s^\alpha+\gamma_\alpha )^m}\right].
\label{eq:disorder14}
\end{align}
We note the useful relation given by 
\cite{Podlubny}
\begin{align}
{\cal L}^{-1} \left[ 
\frac{1}{s^{(2-\alpha)m+1} \left(s^\alpha+\gamma_\alpha\right)^m} 
\right]=\frac{t^{2m}}{(m-1)!} E_{\alpha,(2-\alpha)m+1+\alpha}^{(m-1)}
\left(-\gamma_\alpha t^\alpha \right), 
\label{eq:Pod}
\end{align}
where $E_{a,b}(x)$ denotes the generalized Mittag--Leffler function and 
$E_{a,b}^{(n)}(x)$ is its derivatives of order $n$. 
The generalized Mittag--Leffler function is defined by 
$E_{a,b}(x)=\sum_{n=0}^\infty x^n/\Gamma(an+b)$. 
The inverse Laplace transform of Eq.~(\ref{eq:disorder14})  can be performed 
by using Eq.~(\ref{eq:Pod}). 
The result is given by Eq.~(\ref{eq:disorder15}).

%\newpage
%\renewcommand{\theequation}{B.\arabic{equation}}  
%\setcounter{equation}{0}  % reset counter     
%\section*{Appendix B. Derivation of the upper limit}
%%\section*{Appendix B. }

%\bibliography{ESR_mn_n} 
%

\end{document}